\documentclass[12pt]{article}
\usepackage{pdfpages}
\renewcommand{\theequation}{\arabic{section}.\arabic{equation}}
 
\def\be{\begin{equation}}
\def\ee{\end{equation}}
\def\bqa{\begin{eqnarray}}
\def\eqa{\end{eqnarray}}
\def\lsim{\raise0.3ex\hbox{$<$\kern-0.75em\raise-1.1ex\hbox{$\sim$}}}
\def\gsim{\raise0.3ex\hbox{$>$\kern-0.75em\raise-1.1ex\hbox{$\sim$}}}
\usepackage{amssymb}

\begin{document}

\begin{center}
{\bf Problems with Ultrahigh-energy Neutrino Interactions}\footnote{Presented
at the International School of Subnuclear Physics, 52nd Course, Erice-Sicily,
24 June - 3 July 2014}
\end{center}
\vspace {0.5cm}
\begin{center}
{\bf Dieter Schildknecht\footnote{Email: schild@physik.uni-bielefeld.de}}
\\[2.5mm]
Fakult\"at f\"ur Physik, Universit\"at Bielefeld,\\[1.2mm]
  Universit\"atsstra\ss e 25, 33615 Bielefeld, Germany\\[1.2mm]
and\\[1.2mm]
Max-Planck-Institute for Physics, F\"ohringer Ring 6, \\[1.2mm]
80805 Munich, Germany
\end{center} 

\vspace{1cm}

\begin{center}
{\bf Abstract}
\end{center}
The IceCube collaboration has recently identified events due to
ultrahigh-energy neutrino interactions. Predictions of the neutrino-nucleon
cross section at ultrahigh energies require a huge extrapolation of the
cross sections experimentally measured at laboratory energies. Upon
relating neutrino scattering to deep inelastic electron scattering,
we show that the empirically verified color dipole picture is well suited
for such an extrapolation. The dominant contribution to the total
neutrino-nucleon cross section, even at ultrahigh energies, is due to
the kinematic range where color transparency is valid for the 
color dipole interaction. We deviate from various claims in the literature
on the presence of screening effects due to non-linear evolution at
ultrahigh neutrino energies.

\section{The IceCube Experiment}
\label{}

The IceCube experiment, located at the South Pole, covers a cubic kilometer
of Antarctic glacial ice. It detects neutrinos by observing Cherenkov light
from secondary charged particles produced in neutrino-nucleon interactions.
In 2013, the IceCube collaboration announced the detection of 28 neutrino
events of ultra-high energy:
\begin{itemize}
\item 26 events in the energy range of 50 TeV to 1 PeV = $10^{15}$ eV,
\item 2 events in the energy range of 1 PeV to 2 PeV,
\item zero events above 2 PeV.
\end{itemize}

\begin{figure}[h]
\begin{center}
\includegraphics[scale=.4]{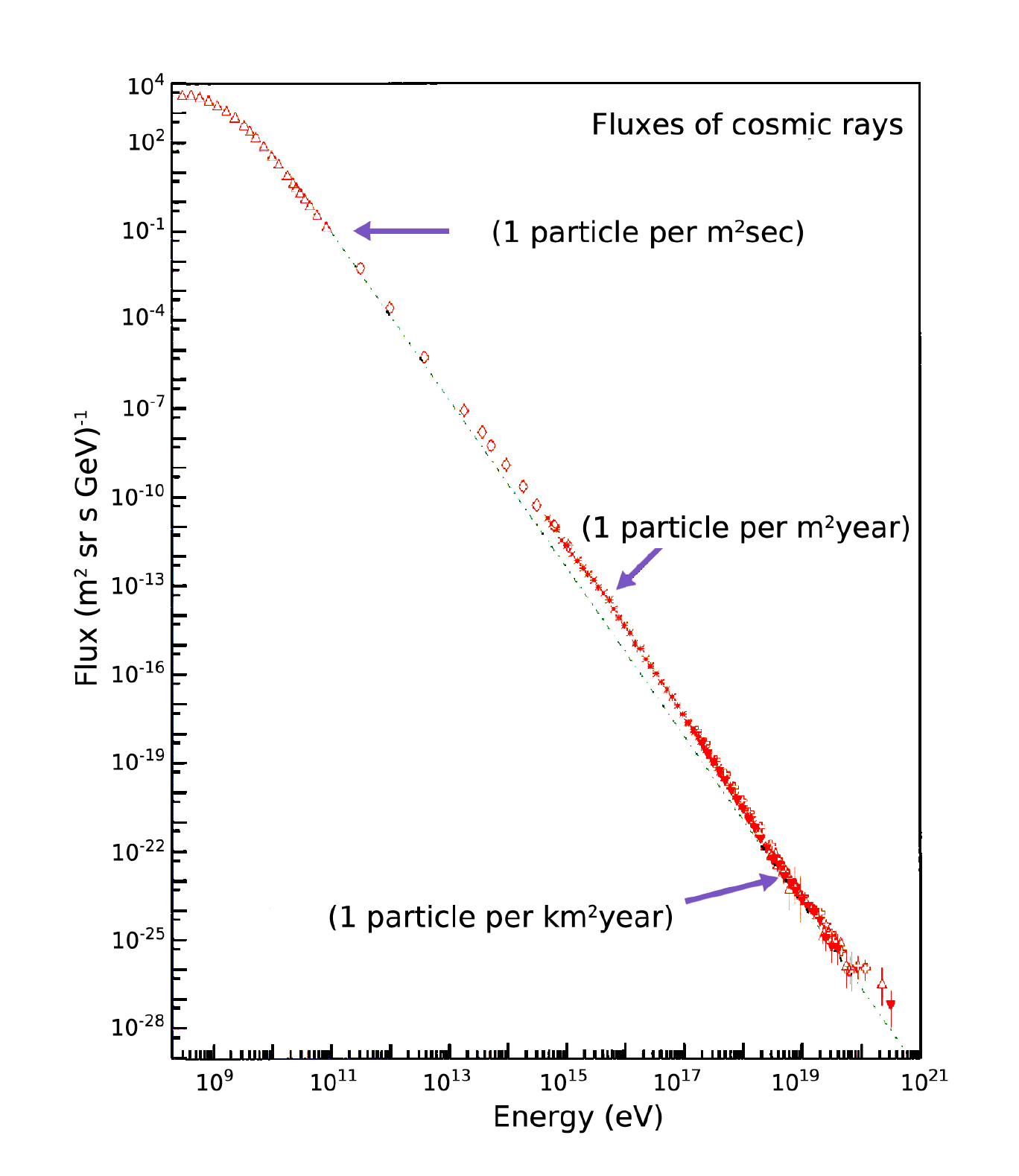}
\vspace*{-0.5cm}
\caption{Flux of cosmic rays as a function of their energy.}
\vspace*{-0.5cm}
\end{center}
\end{figure}

In fig.\,1, I show the flux (e.g. \cite{Bietenholz})
of cosmic rays. The ultrahigh energies of the
observed neutrino events are significantly below the upper end of the
energy spectrum of cosmic rays.

A recent, eighty-page review paper, entitled ``Cosmic Neutrino Pevatrons: A
Brand New Pathway to Astronomy, Astrophysics, and Particle Physics''
\cite{Anchor}, 
contains
a thorough discussion on the interpretation of the IceCube events. It covers 
potential sources of the neutrino events, galactic and extragalactic ones,
as well as probes of fundamental physics. Cosmic probes include e.g.
superheavy Dark Matter, lepto-quarks produced via $\nu_\tau + q \to LQ \to \tau
+ q$, and others.

Let me only mention a single recent speculative remark \cite{Barger}
on the possibility of
an exotic neutrino property. The suggestion is as simple, as it is far
reaching at the same time. The afore-mentioned energies of the IceCube
events (in distinction from the upper end of the cosmic-ray flux in fig.\,1)
do not extend beyond approximately 2 PeV. In particular, the enhancement
of the neutrino signal due to the so-called Glashow resonance \cite{Glashow}
$$
\nu_e e^- \to W^- \to \mbox{anything},
\nonumber
$$
corresponding to an energy of about $E \simeq 6.3$ PeV in the rest
frame of the interacting electron $e^-$, has not been observed (up to now).
Interpreting this non-observation as the consequence of a strict universal
bound on the neutrino energy, $E^\nu_{Max}$, located below 6.3 PeV, from
$$
E^\nu_{Max} = \frac{m_\nu}{\sqrt{1 - \beta^2_\nu}} = \gamma_\nu m_\nu,
$$
one concludes that $\gamma_\nu = E^\nu_{Max}/m_\nu < \infty$ with the
dramatic consequence of a violation of Lorentz invariance for neutrinos.

With respect to future searches for, and for the detailed analysis of
ultrahigh-energy neutrino events, a knowledge of the neutrino-nucleon
cross section is indispensable. Predictions of the cross section for
ultrahigh energies of neutrinos require huge extrapolations to energies
far beyond the energies where experimental data from laboratory experiments
are available. The highest energies in laboratory experiments on lepton
scattering were reached for charged-lepton scattering at the 
electron-proton collider HERA in Hamburg, which was running during the 
years from 1990 to 2007. By relating neutrino scattering to charged-lepton
scattering, the HERA experimental data can be used as a reliable base for
the extrapolation to the ultrahigh energies observed for cosmic neutrinos.

\section{The Neutrino-Nucleon Cross Section at Ultrahigh Energies.}
\renewcommand{\theequation}{\arabic{section}.\arabic{equation}}
\setcounter{equation}{0}

\subsection{Connection to Deep Inelastic Electron-Nucleon Scattering.}

The total charged-current neutrino-nucleon cross section is given by
(e.g. \cite{Goncalves})
\be
\sigma_{\nu N} (E) = \frac{G^2_F}{2 \pi} \int^{s-M_p^2}_{Q^2_{min.}} dQ^2
\left( \frac{M^2_W}{Q^2 + M^2_W} \right)^2 \int^{s-Q^2}_{M_p^2} 
\frac{dW^2}{W^2} \sigma_r (x,Q^2).
\label{2.1}
\ee
In (\ref{2.1}), $E$ denotes the neutrino energy in the nucleon rest frame.
It is related to the neutrino-nucleon center-of-mass energy by
\be
s = 2 M_pE + M_p^2 \cong 2 M_pE,
\label{2.2}
\ee
where $M_p$ denotes the proton mass. The integrations in (\ref{2.1})
run over the four-momentum-transfer squared from the neutrino to the
nucleon, $Q^2$, and the square of the total energy, $W^2$, of the produced
hadronic system. The Fermi coupling is denoted by $G_F$, and the
W-boson mass by $M_W$. The reduced cross section, $\sigma_r(x,Q^2)$ in
(\ref{2.1}), depends on the three nucleon structure functions
$F^{\nu N}_{2,L,3} (x, Q^2)$,
\be
\sigma_r (x,Q^2) = \frac{1+(1-y)^2}{2} F_2^{\nu N} (x,Q^2) - \frac{y^2}{2}
F_L^{\nu N} (x,Q^2) + y (1-\frac{y}{2}) x F_3^{\nu N} (x,Q^2),
\label{2.3}
\ee
where $x$ is the Bjorken variable,
\be
x = \frac{Q^2}{2qP} = \frac{Q^2}{W^2+Q^2-M_p^2} \cong \frac{Q^2}{W^2},
\label{2.4}
\ee 
and $y$ denotes the squared fraction of the total neutrino energy producing
hadrons in the final state 
\be
y = \frac{Q^2}{2M_pEx} \cong \frac{W^2}{s}.
\label{2.5}
\ee
According to (\ref{2.1}), at ultra-high energies, $s\gg M^2_W$, the 
contribution to the total cross section due to $Q^2 \gg M^2_W$ is
strongly suppressed. The dominant contribution to the cross section is
due to $Q^2 \cong M^2_W$ and due to small values of
$x \cong M^2_W/s \ll 0.1.$

At low values of $x \lsim 0.1$, neutrino-nucleon and electron-nucleon
interactions proceed via fluctuations of the virtual W-boson and virtual
photon, respectively, into quark-antiquark pairs that propagate and interact
via gluon exchange with the nucleon. For a given number of $n_f$ actively
contributing quark flavors, the flavor-independent QCD interaction implies
the proportionality
\be
\frac{1}{n_f} F^{\nu N}_{2,L} (x , Q^2) = 
\frac{1}{\sum_q Q^2_q} F^{eN}_{2,L} (x , Q^2).
\label{2.6}
\ee
It allows to predict the neutrino-nucleon structure functions, 
$F^{\nu N}_{2,L} (x,Q^2)$, 
from the electron-nucleon structure functions, $F^{eN}_{2,L} (x,Q^2)$.
For $n_f = 4$ flavors, the proportionality factor in (\ref{2.6}) becomes
$n_f/\sum^{n_f}_q Q^2_q = 5/18$.

The dominant term in (\ref{2.3}) is due to $F^{\nu N}_2 (x, Q^2)$. In what
follows, we shall frequently use the relation between the proton
structure function, $F_2^{ep} (x, Q^2)$, and the (virtual) photoabsorption
cross section, $\sigma_{\gamma^* p} (W^2,Q^2)$, that, for $x \ll 0.1$, is
given by
\be
F^{ep}_2 (x,Q^2) = \frac{Q^2}{4 \pi^2 \alpha} \sigma_{\gamma^*p} (W^2,Q^2).
\label{2.7}
\ee

\subsection{Electron-Proton Deep Inelastic Scattering, Experimental Results.}

As noted, neutrino scattering at ultrahigh energies is dominantly due
to interactions in the kinematic range of $x \ll 0.1$. This is the
region of the kinematic variables $Q^2$ and $W^2$, where the HERA
experimental results show scaling in the low-x scaling 
variable \cite{DIFF2000},
\be
\eta(W^2,Q^2) \equiv \frac{Q^2 + m^2_0}{\Lambda^2_{sat} (W^2)}, 
\hspace*{2cm}
\Lambda^2_{sat} (W^2) \sim (W^2)^{C_2}
\label{2.8}
\ee
i.e.
\be
\sigma_{\gamma^*p} (W^2, Q^2) =  \sigma_{\gamma^*p} (\eta (W^2,
Q^2))  \sim  \sigma^{(\infty)} \left\{ \begin{array}{l@{\quad,\quad}l}
\frac{1}{\eta (W^2, Q^2)} & ~{\rm for}~ \eta (W^2, Q^2) \gg 1 , \\
\ln \frac{1}{\eta (W^2,Q^2)} & ~{\rm for}~ \eta(W^2,Q^2) \ll 1 .
\end{array} \right.
\label{2.9}
\ee
The photoabsorption cross section, in the approximation where the
(hadronic) cross section $\sigma^{(\infty)} (W^2) = \sigma^{(\infty)} \cong
const$, depends on the single scaling variable $\eta (W^2, Q^2)$ defined
in (\ref{2.8}). Compare fig. 2.

\begin{figure}[h!]
\begin{center}
\includegraphics[scale=.4]{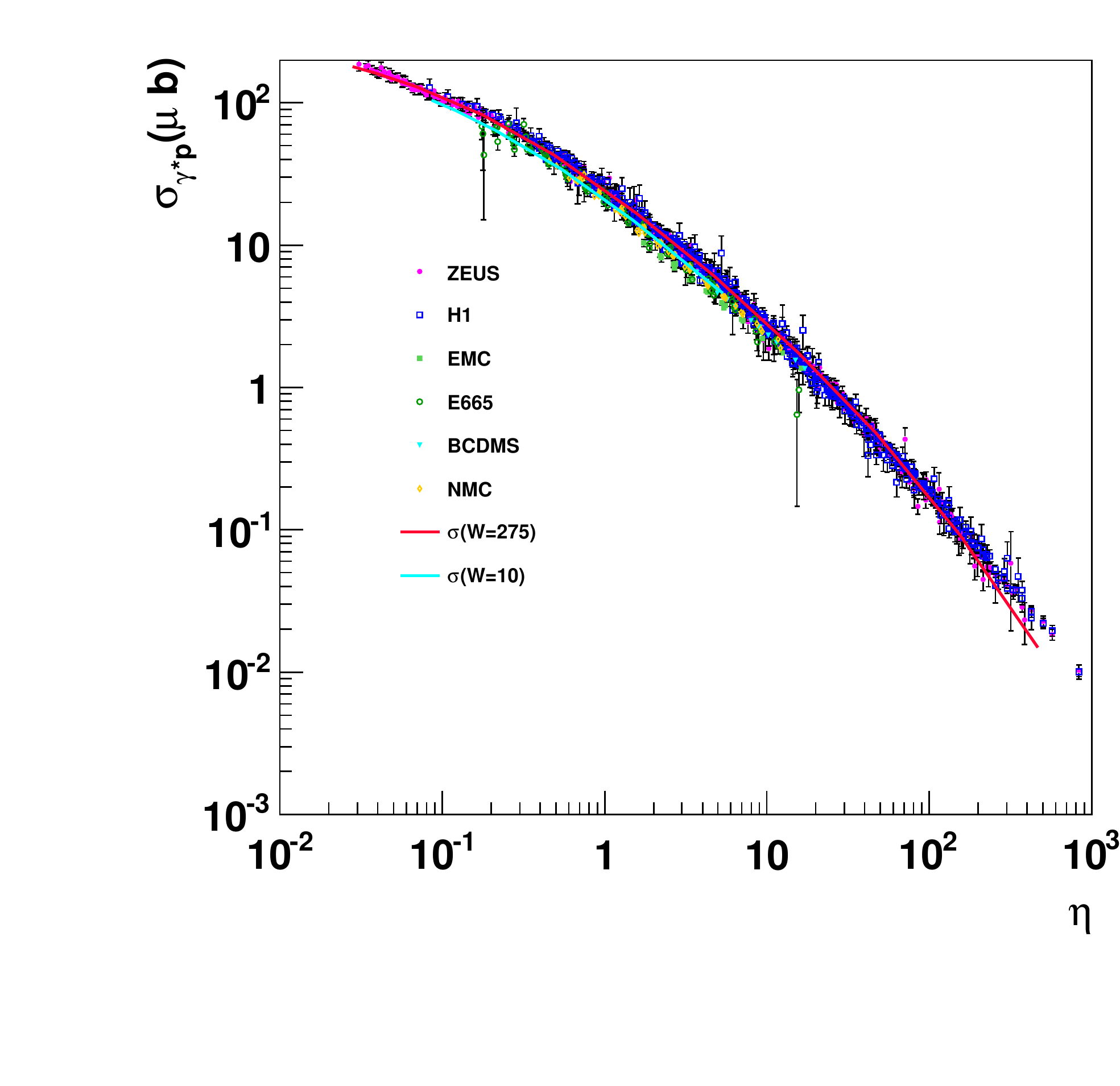}
\vspace*{-1.3cm}
\caption{The total photoabsorption cross section as a function of the
low-x scaling variable \cite{DIFF2000} $\eta \equiv \eta (W^2,Q^2)$.}
\vspace*{-0.3cm}
\end{center}
\end{figure}

The ``saturation scale'', $\Lambda^2_{sat} (W^2)$, grows as 
$\Lambda^2_{sat} (W^2) \sim (W^2)^{C_2}$, where $C_2 \simeq 0.3$, and
the constant $m_0$, as expected from quark-hadron duality 
\cite{ST} for light
quarks, is somewhat smaller than the $\rho^0$-meson mass, $m_{\rho^0}$.
The functional dependence (\ref{2.9}) of the photoabsorption cross section
on $\eta (W^2,Q^2)$ can be read off from fig. 2.

We remark that the theoretical curves in fig.\,2 were obtained in the 
color-dipole picture (CDP) to be discussed in the next section. The slight
spread in the predictions for $W = 275$ GeV and $W = 10$ GeV, also visible
in the experimental data, is due to the deviation of $\sigma^{(\infty)}
(W^2)$ in (\ref{2.9}) from $\sigma^{(\infty)} (W^2) = \sigma^{(\infty)} =
\mbox{const}$.

The significant differences in the $\eta (W^2, Q^2)$ dependence for
$\eta (W^2,Q^2) \gg 1$ and $\eta (W^2,Q^2) \ll 1$ in fig.\,2 suggest a
subdivision of the $(Q^2, W^2)$ plane into (only and precisely) two regions
subdivided by $\eta (W^2,Q^2) = 1$, compare fig. 3.
\begin{figure}[h]
\vspace*{-1cm}
\begin{center}
\includegraphics[scale=.65]{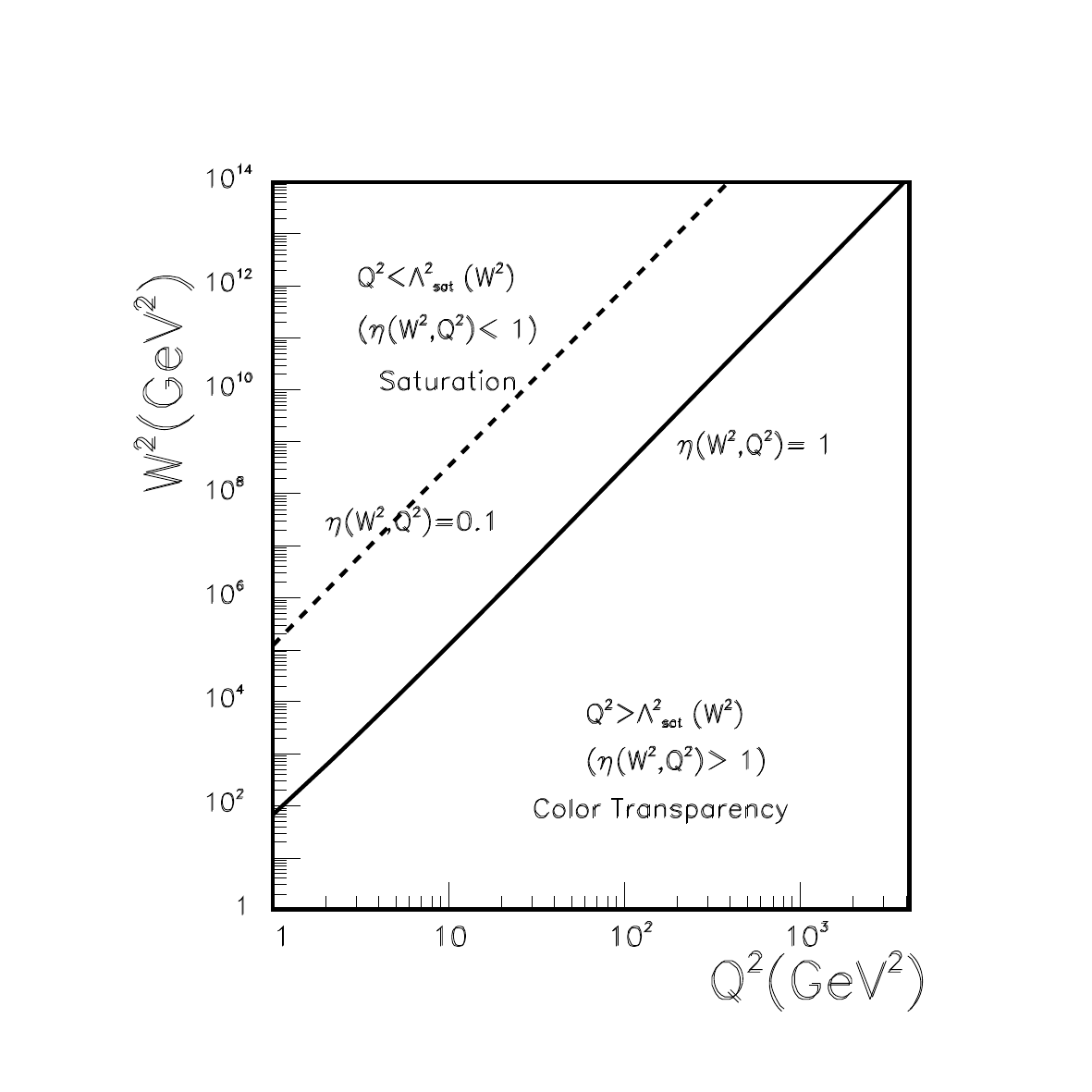}
\vspace*{-0.5cm}
\caption{The $(Q^2,W^2)$ plane with the line $\eta(W^2,Q^2) = 1$ separating
the large-$Q^2$ color-transparency region from the saturation domain.}
\end{center}
\end{figure}

The logarithmic dependence of $\sigma_{\gamma^*p} (\eta (W^2,Q^2))$ for
$\eta (W^2,Q^2) < 1$ implies the important limiting behavior 
\cite{DIFF2000, SCHI},
\bqa
\lim_{W^2 \to \infty \atop {Q^2~{\rm fixed}}} 
\frac{\sigma_{\gamma^*p} (\eta (W^2, Q^2))}{\sigma_{\gamma^*p} (\eta
(W^2, Q^2 = 0))} 
 & = &\lim_{W^2 \to \infty \atop {Q^2~{\rm fixed}}} 
\frac{\ln \left( \frac{\Lambda^2_{sat}(W^2)}{m^2_0} 
\frac{m^2_0}{(Q^2 + m^2_0)} \right)}{\ln 
\frac{\Lambda^2_{sat} (W^2)}{m^2_0}} \nonumber \\
& = & 1 + \lim_{W^2 \to \infty \atop {Q^2~{\rm fixed}}} 
\frac{\ln \frac{m^2_0}{Q^2 + m^2_0}}{\ln 
\frac{\Lambda^2_{sat} (W^2)}{m^2_0}} = 1.
\label{2.10}
\eqa
For $W^2 \to \infty$, and any fixed value of the photon virtuality, $Q^2$, the
ratio of the photoabsorption cross section, $\sigma_{\gamma^*p} (\eta
(W^2,Q^2))$, to the $Q^2 = 0$ photoproduction cross section $\sigma_{\gamma p}
(W^2) \equiv \sigma_{\gamma^* p} (\eta (W^2,Q^2 = 0))$, becomes equal
to unity. The photoabsorption cross section ``saturates'' to a unique
limit that coincides with the hadronlike $Q^2 = 0$ photoproduction cross
section.

The scaling behavior of $\sigma_{\gamma^*p} (\eta (W^2,Q^2))\sim 1/\eta
(W^2,Q^2)$ for $\eta (W^2,Q^2) \gg 1$, in terms of the proton structure
function $F_2 \equiv F^{ep}_2(x,Q^2)$ at sufficiently large $Q^2$, corresponds to a
single curve in the plot of the experimental data for $F_2$
against the variable $1/W^2$. The theoretical curve in fig. 4
\begin{figure}[h]
\begin{center}
\includegraphics[scale=.55]{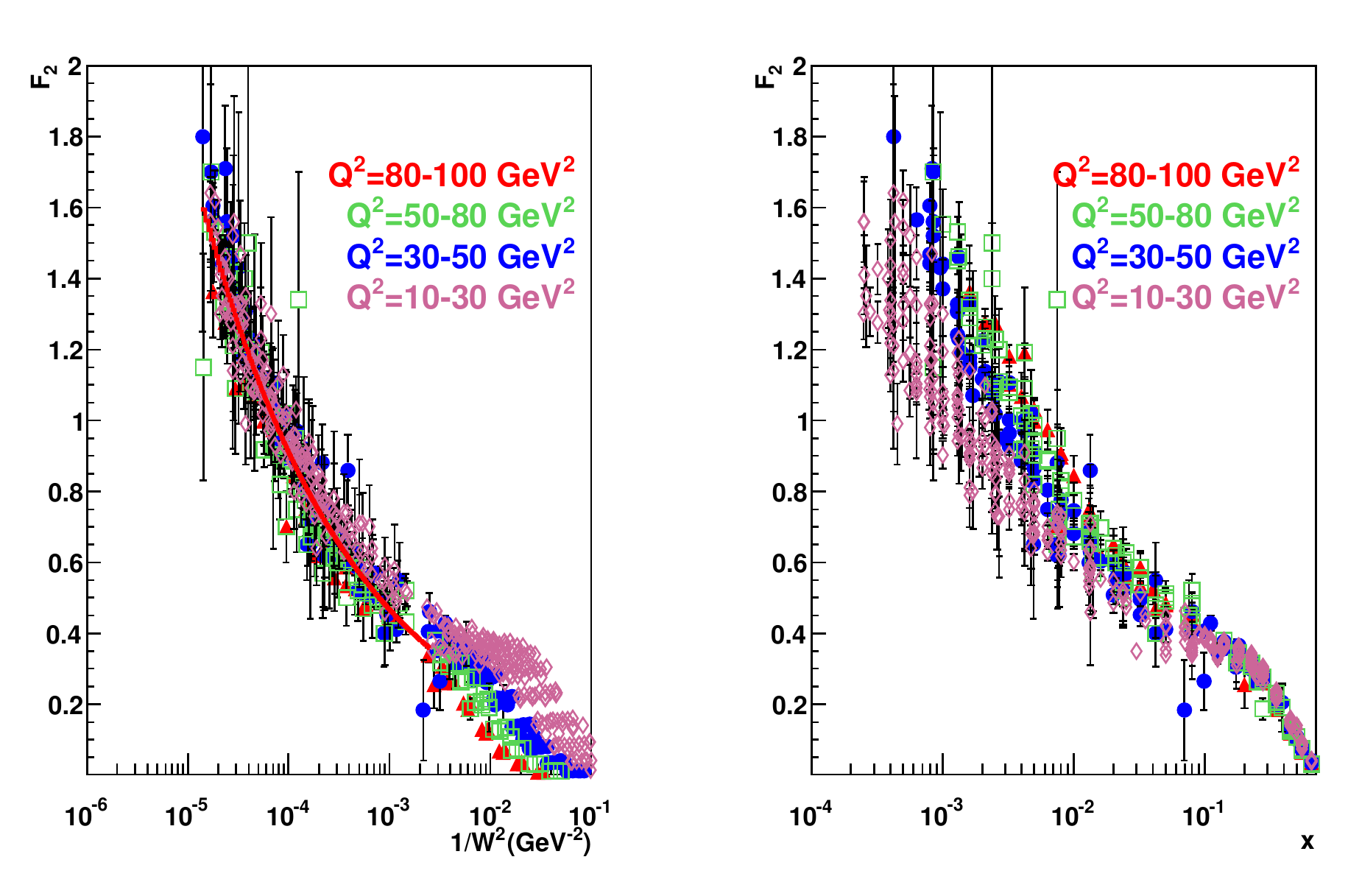}
\vspace*{-0.5cm}
\caption{Experimental results for the proton structure function
$F_2 \equiv F^{ep}_2 (x, Q^2)$ as a function of $1/W^2$ and as a function 
of $1/x$ \cite{E}.}
\vspace*{-1cm}
\end{center}
\end{figure}
\begin{figure}[h!]
\begin{center}
\includegraphics[scale=.4]{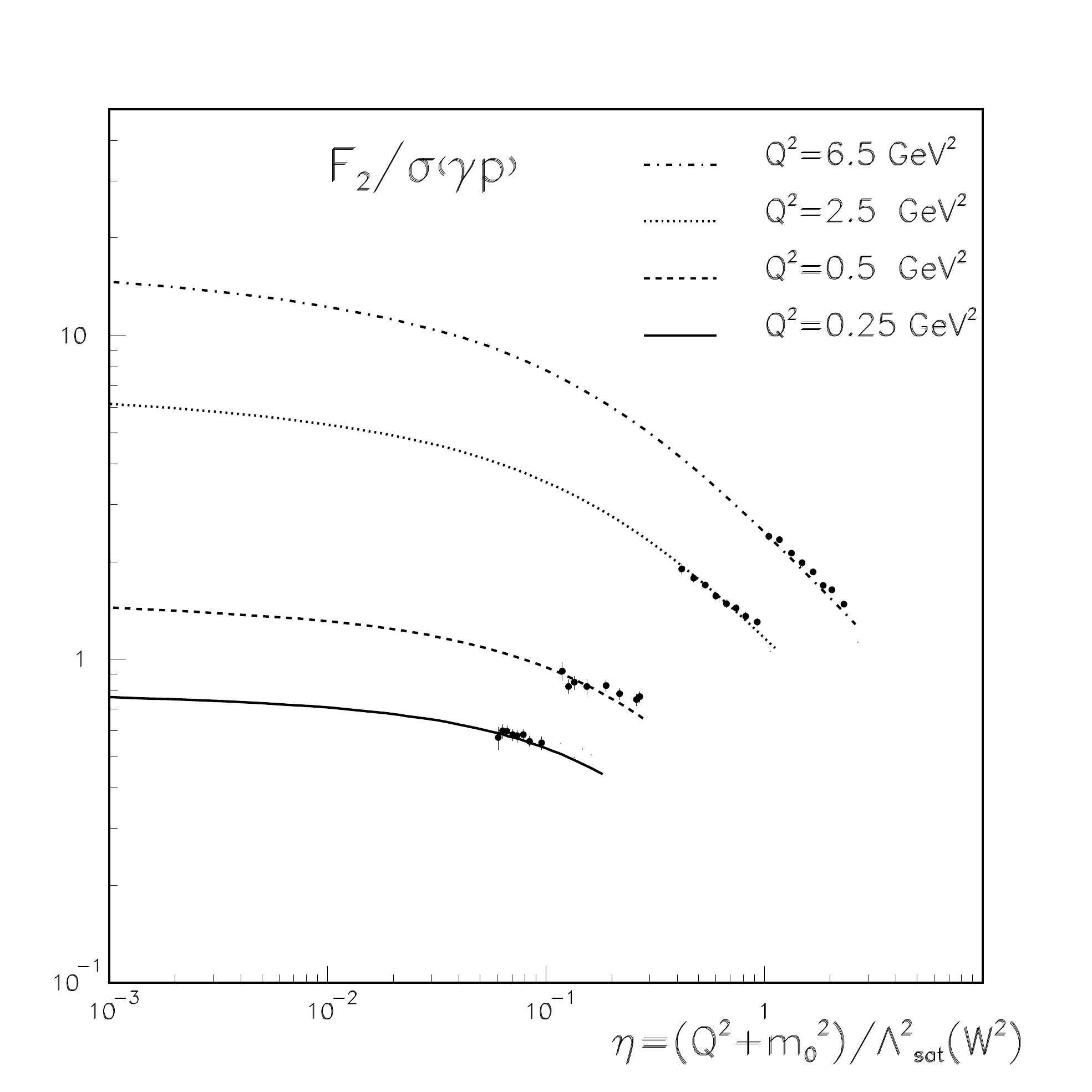}
\vspace*{-0.5cm}
\caption{The approach to the saturation limit \cite{E}. With decreasing
$\eta (W^2,Q^2)$, the proton structure function $F_2 (x,Q^2)$ approaches
the saturation limit given by $Q^2 \sigma_{\gamma p} (W^2)$, where
$\sigma_{\gamma p} (W^2)$ denotes the $(Q^2 = 0)$ photoabsorption cross
section.}
\end{center}
\end{figure}
is based on \cite{E}
\be
F_2 (W^2) = f_2 \cdot \left( \frac{W^2}{1{\rm GeV}^2} \right)^{C_2 = 0.29},
\label{2.11}
\ee
where $f_2 = 0.063$.
For comparison, in fig. 4, we also show $F_2$ as a function of the
Bjorken variable $x$ for fixed values of $Q^2$. 

The transition to the saturation limit
for $\eta (W^2,Q^2) \ll 1$, based on the CDP \cite{E}
is illustrated in fig. 5.
Even for values of $Q^2$ as small as $Q^2 \cong 1 \mbox{GeV}^2$,
a strict and detailed empirical verification of the saturation limit
requires ``ultrahigh'' energies far beyond the energies that were
available at HERA.

The afore-mentioned experimental results on low-x DIS find a coherent
{\it qualitative as well as quantitative} explanation in the color
dipole picture (CDP) to which we shall turn next.

\subsection{The Color Dipole Picture: Theory}

In DIS at low $x \lsim 0.1$, the photon virtually dissociates (or
``fluctuates'' in modern jargon) into hadronic vector states that
subsequently interact with the nucleon (generalized vector dominance
(GVD) \cite{Sakurai}, \cite{PP2005}). In QCD, the hadronic vector states are 
quark-antiquark $(q \bar q)$
color-dipole states, which interact with the gluon field in the nucleon.
Compare fig. 6 for the imaginary part of the associated
forward-Compton-scattering amplitude.
\begin{figure}[h]
\begin{center}
\includegraphics[scale=.8]{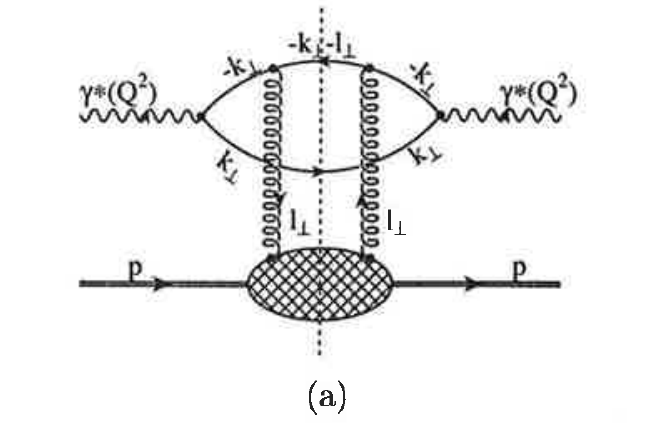}\hspace*{1cm}
\includegraphics[scale=.8]{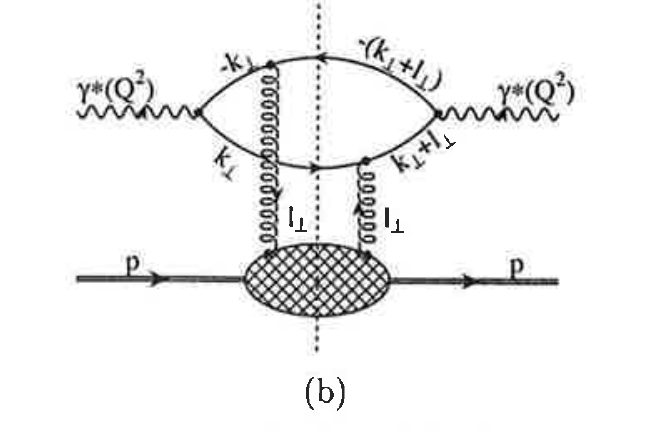}
\vspace*{-0.3cm}
\caption{Two of the four diagrams for the $q \bar q$ dipole interaction with
the gluon field in the nucleon. The diagrams (a) and (b) correspond to
channel 1 and channel 2 respectively.}
\end{center}
\vspace*{-0.6cm}
\end{figure}

In terms of the transverse size, $\vec r_\bot$, of the $q \bar q$ pair,
and the partition $0 \le z \le 1$ of the longitudinal momenta of quark
and antiquark, the photoabsorption cross section, $\sigma_{\gamma^*_{L,T} p}
(W^2,Q^2)$, for longitudinally and transversely polarized photons takes the
form \cite{Nikolaev, Cvetic}
\bqa
\sigma_{\gamma^*_{L,T}p} (W^2, Q^2) & = &
\int dz \int d^2 \vec r_\bot
\vert \psi_{L,T} (\vec r_\bot, z (1 - z), Q^2) \vert^2 \nonumber \\
& \times & \sigma_{(q\bar q)p} (\vec r_\bot, z (1 - z), W^2). 
\label{2.12}
\eqa
The integrand in the photoabsorption cross section (\ref{2.12}) shows
the afore-mentioned factorization into the $Q^2$-dependent probability
for a $\gamma^* \to q \bar q$ fluctuation and the W-dependent $(q \bar q)p$
interaction cross section. 

The gauge-invariant two-gluon interaction of
the $q \bar q$ color-dipole state requires the representation \cite{Nikolaev,
Cvetic}
\be
\sigma_{(q \bar q)p} (\vec r_\bot, z(1-z),W^2) 
=  \int^{\vec l^{~2}_{\bot~Max} (W^2)}
d^2 \vec l_\bot \tilde \sigma \left( \vec l^{~2}_\bot, z (1-z), W^2 \right) 
\left( 1-e^{-i \vec l_\bot \vec r_\bot}\right).
\label{2.13}
\ee
The CDP, accordingly, rests on the general form (\ref{2.12}) of the 
photoproduction cross section at low $x$, combined with the 
constraint on the dipole cross section (\ref{2.13})\footnote{We stress
the dependence of the dipole cross section (\ref{2.13}) on the energy
$W$, \cite{Sakurai}, \cite{Cvetic}, \cite{DIFF2000}, \cite{PP2005},
\cite{Ewerz}
as implied by the fluctuation of the photon into on-shell $q \bar q$
dipole states, forbidding the $Q^2$ dependence (via Bjorken $x \sim Q^2/W^2$
dependence) frequently assumed in the literature.}

Employing the explicit (QED) expression for the $\gamma^* \to q \bar q$
transition probability, the photoabsorption cross section (\ref{2.12})
may be written as \cite{E}
\be
\sigma_{\gamma^*_{L,T}p} (W^2, Q^2) = 
\frac{\alpha}{\pi} \sum_q Q^2_q Q^2 
\int dr^{\prime 2}_\bot K^2_{0,1} (r^\prime_\bot Q) 
\sigma_{(q \bar q)^{J=1}_{L,T} p} (r^{\prime 2}_\bot, W^2).
\label{2.14}
\ee
In the transition from (\ref{2.12}) to (\ref{2.14}), massless quarks
were assumed, and, by the appropriate projection, the cross sections
$\sigma_{(q \bar q)_{L,T}^{J=1}p} (r^{\prime 2}_\bot, W^2)$ for the
scattering of $(q \bar q)^{J=1}_{L,T}$ states were introduced. The sum
$\sum_q Q^2_q$ in (\ref{2.14}) runs over the actively contributing
quark flavors, and $K_0 (r^\prime_\bot Q)$ and $K_1 (r^\prime_\bot Q)$ 
denote modified 
Bessel functions. In terms of the $J=1$ projections, 
$\sigma_{(q \bar q)^{J=1}_{L,T}p} (r^\prime_\bot, W^2)$, of the dipole
cross section in (\ref{2.12}), the two-gluon structure of the dipole
cross section from (\ref{2.13}) becomes \cite{E}
\be
\sigma_{(q \bar q)^{J=1}_{L,T}p} (r^\prime_\bot, W^2)  =
\pi \int d \vec l^{~\prime 2}_\bot \bar \sigma_{(q \bar q)^{J=1}_{L,T} p}
(\vec l^{~\prime 2}_\bot , W^2)
\left( 1 - \frac{\int d \vec l^{~\prime 2}_\bot 
\bar \sigma_{(q \bar q)^{J=1}_{L,T} p} (\vec l^{~\prime 2}_\bot, W^2) J_0
(l^\prime_\bot r^\prime_\bot)}{\int d \vec l^{~\prime 2}_\bot
\bar \sigma_{(q \bar q)^{J=1}_{L,T} p} (\vec l^{~\prime 2}_\bot, W^2)}
\right), 
\label{2.15}
\ee
where $J_0 (l^\prime_\bot r^\prime_\bot)$ is the Bessel function with
index 0.

The representation of the photoabsorption cross section (\ref{2.14}) together
with the constraint (\ref{2.15}) allows one to {\it derive} 
\cite{E}, \cite{Erice} the
important {\it qualitative} behavior of the photoabsorption cross section
(\ref{2.9}) seen in the experimental data in fig. 2.

For any fixed value of $Q^2$, the strong decrease of the functions
$K^2_{0,1} (r^\prime_\bot Q)$ restricts the range of $r^\prime_\bot$
relevant for the integration over the dipole size $r^{\prime2}_\bot$
in (\ref{2.14}). The behavior of the dipole cross section in (\ref{2.14})
for this $Q^2$-dependent  effective dipole size $r^{\prime 2}_\bot$, limited
according to (\ref{2.15}), is determined by the interaction energy, $W$,
of the $q \bar q$ dipole state. Assuming that the effective upper limit of
the integration over $d \vec l^{~\prime 2}_\bot$ in (\ref{2.15}), given
by $\vec l^{~\prime 2}_{\bot Max} (W^2)$, increases with the energy $W$,
two different limiting behaviors of the dipole cross section at a fixed
value of $r^\prime_\bot$ can be discriminated.

For the case of relatively low energy, namely for $0 < l^\prime_\bot
r^\prime_\bot < l^\prime_{\bot Max} (W^2) r^\prime_\bot \ll 1$, 
the expansion of
the Bessel function in (\ref{2.15}) implies that the dipole-proton
cross section (\ref{2.15}) vanishes proportional to the dipole size
squared (``color transparency'' limit),
\be
\sigma_{(q \bar q)^{J=1}_{L,T}p} (r^{\prime 2}_\bot , W^2) \sim
r^{\prime 2}_\bot.
\label{2.16}
\ee
The proportionality factor in (\ref{2.16}) is given by the first moment
of $\bar \sigma_{(q \bar q)^{J=1}_{L,T}p} (\vec l^{~\prime 2}_\bot, W^2)$,
and is identified with the saturation scale $\Lambda^2_{sat} (W^2)$ 
introduced in (\ref{2.8}).

For the case of very high energy, such that $l^{\prime 2}_{\bot Max} (W^2)
r^\prime \gg 1$, rapid oscillations of the Bessel function $J_0 
(l^\prime_\bot r^\prime_\bot)$ in (\ref{2.15}) imply (``saturation''
limit).
\be
\lim_{r^{\prime 2} fixed \atop W^2 \to \infty} 
\sigma_{(q \bar q)^{J=1}_{L,T} p} (r_\bot^{~\prime 2}, W^2) 
=
\lim_{r^{\prime 2}_\bot \to \infty \atop W^2 = const} 
\sigma_{(q \bar q)^{J=1}_{L,T}p} (r^{\prime 2}_\bot, W^2) = 
\sigma_{L,T}^{(\infty)} (W^2).
\label{2.17}
\ee
At fixed dipole size, for $W^2$ sufficiently large the dipole cross
section converges to the hadronic cross section that at most depends
weakly on $W^2$, i.e. $\sigma^{(\infty)}_{L,T} (W^2) \simeq \sigma^{(\infty)}
= \mbox{const}$.

Evaluating the photoabsorption cross section (\ref{2.12}) for the
two different limits, (\ref{2.16}) and (\ref{2.17}), we find \cite{E, Erice}
\bqa
\hspace*{-0.5cm}
\sigma_{\gamma^*p} (W^2,Q^2) & = & \sigma_{\gamma^*p} 
(\eta (W^2,Q^2)) \nonumber \\
 & =  & \frac{\alpha}{\pi} \sum_q Q^2_q
\left\{
  \begin{array}{l@{\quad,\quad}l}
\frac{1}{6} (1 + 2 \rho) \sigma_L^{(\infty)} (W^2) \frac{1}{\eta (W^2,Q^2)} 
& (\eta (W^2,Q^2) \gg 1) \\
\sigma_T^{(\infty)} (W^2) \ln \frac{1}{\eta (W^2,Q^2)} 
& (\eta (W^2,Q^2) \ll 1)
\end{array} \right.
\label{2.18}
\eqa
The form (\ref{2.18}) agrees with the empirical result 
(\ref{2.9})\footnote{The quantity $\rho$ yields the longitudinal-to-transverse
ratio of the respective photoabsorption cross sections, $R=1/2 \rho$, for
$\eta (W^2,Q^2) \gg 1$. The ratio $\rho \gsim 4/3$  \cite{E} gives
the relative transverse sizes of transversely-versus-longitudinally polarized
$q \bar q$ dipole states in the color-transparency
limit (\ref{2.16}).}. 
The
form of the $q \bar q$ color-dipole interaction (\ref{2.13}) dictated by gauge
invariance, {\it implies scaling} in $\eta (W^2,Q^2)$ of the total
photoabsorption cross section. The empirical functional dependence,
as $1/\eta (W^2,Q^2)$ and as $\ln (1/\eta (W^2,Q^2))$ in (\ref{2.9}),
is traced back to {\it color transparency and saturation} of the $(q \bar q) p$
color dipole interaction.

\subsection{ A Remark on the Gluon Distribution.}

The proton structure function in the CDP, according to (\ref{2.18}) as well
as (\ref{2.11}) and\break fig. 4, is given by
\be
F_2^{ep} (x,Q^2) \sim 
\left\{
  \begin{array}{l@{\quad , \quad}l}
\sigma^{(\infty)} \Lambda^2_{sat} (W^2) 
& (\eta (W^2,Q^2) \gg 1) \\
Q^2 \sigma^{(\infty)} \ln \frac{\Lambda^2_{sat} (W^2)}{Q^2 + m^2_0}
& (\eta (W^2,Q^2) \ll 1)
\end{array} \right.
\label{2.19}
\ee
Noting that in the perturbative-QCD-improved parton model the
structure function $F_2^{ep} (x, Q^2)$ is proportional to the gluon-distribution
function, we have to identify \cite{E}
\be
\sigma^{(\infty)} (W^2) \Lambda^2_{sat} (W^2) \sim \alpha_s (Q^2) x g 
(x,Q^2) \vert_{x = \frac{Q^2}{W^2}}.
\label{2.20}
\ee
The $W$ dependence of the structure function, $F_2^{ep} (x,Q^2)$ at large
values of $\eta (W^2,Q^2) \gg 1$, compare (\ref{2.19}) and
fig. 4, implies that the
gluon structure function depends on the single variable $W^2$. Since the
CDP uniquely fixes the $Q^2$ dependence of $F_2^{ep} (x,Q^2)$
for $\eta (W^2,Q^2) \ll 1$, a 
measurement of $\Lambda^2_{sat} (W^2)$ for $\eta (W^2,Q^2) \gg 1$
determines the structure function in the saturation limit of
$\eta (W^2,Q^2) \ll 1$, compare the $(Q^2,W^2)$ plane in fig. 3. 

With respect to saturation, our approach of the CDP differs significantly from
the point of view frequently put forward in the literature: therein, the
proportionality of the structure function $F_2 (x, Q^2) \sim \alpha_s (Q^2)
x g(x,Q^2)$ is {\it assumed to persist} in the limit of $W^2 \to \infty$
with $Q^2 = \mbox{const}$, or $x \to 0$, leading to an infinite rise of
$F_2 (x, Q^2)$ for $x \to 0$ at $Q^2$ fixed. Non-linear effects in the
gluon-distribution function \cite{Balitzki}, \cite{Kuo}
associated with quark splitting and 
recombination, must then be invoked to obtain agreement with experiment.

Such an approach ignores the long life time, or the long longitudinal distance
involved in the interaction of a $\gamma^* \to q \bar q$ fluctuation
correctly contained in the CDP. Saturation according to the CDP
corresponds to a transition
from a proportionality to $\Lambda^2_{sat} (W^2)$ to
a logarithmic dependence on this one and the same
function, $\Lambda^2_{sat} (W^2)$, when passing from $\eta (W^2, Q^2) \gg 1$
to $\eta (W^2,Q^2) \ll 1$. This transition is traced back to the different 
interplay, for $\mu (W^2,Q^2) \gg 1$ and $\mu (W^2,Q^2) \ll 1$, between the two
interaction paths of fig. 6 corresponding to the color-gauge-invariant dipole
interaction (\ref{2.13}).   

\subsection{Color Dipole Picture: Quantitative Representation of the
Photoabsorption Cross Section.}

The CDP yields a remarkably simple form for the photoabsorption cross
section, essentially an interpolation between the general
$1/\eta (W^2,Q^2)$ and $\ln (1/\eta (W^2,Q^2))$ dependences given in
(\ref{2.18}). For details we refer to refs. \cite{DIFF2000, E}.
The closed explicit
expression (for $\rho = 1$) reads
\bqa
\sigma_{\gamma^*p} (W^2,Q^2) & = & \frac{\alpha R_{e^+e^-}}{3 \pi}
\sigma^{(\infty)} (W^2) I_0 (\eta (W^2,Q^2)) \nonumber \\ 
&~& \nonumber\\
& \times & \frac{\frac{\xi}{\eta(W^2,Q^2)}}{1 + \frac{\xi}{\eta(W^2,Q^2)}}
 + O \left(
  \frac{m^2_0}{\Lambda^2_{\rm sat} (W^2)} \right)
\label{2.21}
\eqa
where
\be
\frac{\frac{\xi}{\eta(W^2,Q^2)}}{1 + \frac{\xi}{\eta(W^2,Q^2)}} \cong \left\{ \matrix{ 
1 , & {\rm for} ~ \eta (W^2 , Q^2) \ll \xi = 130 \cr 
\frac{\xi}{\eta (W^2 , Q^2)} , & {\rm for} ~\eta (W^2 , Q^2) \gg \xi = 130
}\right.
\label{2.22}
\ee
and $I_0 (\eta (W^2, Q^2))$ is given by
\bqa
\hspace*{-0.5cm}
I_0 (\eta (W^2, Q^2)) & = & 
\frac{1}{\sqrt{1 + 4\eta (W^2, Q^2)}} \ln \frac{\sqrt{1 + 4 \eta (W^2, Q^2)}
  +1}{\sqrt{1+4\eta(W^2, Q^2)}-1} \cong \nonumber \\
& & \nonumber \\
& & \nonumber \\
& \cong& \left\{  \matrix{  \ln \frac{1}{\eta(W^2, Q^2)} + O (\eta \ln \eta ), 
~~~~~~{\rm for}~ \eta (W^2, Q^2) \rightarrow \frac{m^2_0}{\Lambda^2_{\rm sat}
  (W^2)}, \cr
\frac{1}{2\eta (W^2, Q^2)} + O \left( \frac{1}{\eta^2}\right) , ~~~~~~ {\rm
  for}~~ \eta (W^2 , Q^2) \rightarrow \infty   } \right. . 
\label{2.23}
\eqa
The hadronic dipole-proton cross section, $\sigma^{(\infty)} (W^2)$, in
(\ref{2.21})
is determined by requiring consistency of (\ref{2.21})
 with $Q^2 = 0$ photoproduction.
\be
\sigma^{(\infty)} (W^2) = \frac{3\pi}{\alpha R_{e^+e^-}}
\frac{1}{\ln \frac{\Lambda^2_{sat} (W^2)}{m^2_0}} 
\left\{ \begin{array}{l@{\quad\quad}l}
\sigma^{Regge}_{\gamma p} (W^2), & ~ \\
&~ \\
\sigma^{PDG}_{\gamma p} (W^2). & ~
\end{array} \right. .
\label{2.24}
\ee
In (\ref{2.24}), $\sigma^{Regge}_{\gamma p} (W^2)$ asymptotically grows
as a small power of $W^2$ in distinction from $\sigma^{PDG}_{\gamma p}
(W^2)$ that grows as $(\ln W^2)^2$. For the explicit expressions for these
photoproduction cross sections, we refer to refs. \cite{E, PRD88}.

The parameter $\xi$, empirically determined as $\xi \cong 130$, restricts the
masses of $q \bar q$ fluctuations, $M_{q \bar q}$, via $M^2_{q \bar q} \le
m^2_1 (W^2) = \xi \Lambda^2_{sat} (W^2)$. With $m^2_0 = 0.15 GeV^2,
\Lambda^2_{sat} (W^2) = C_1 (W^2/ 1 GeV^2)^{C_2}$, where $C_1 = 0.34 GeV^2$ and
$C_2 \cong 0.27$ to 0.29, one obtains the theoretical results shown in fig. 2.

\subsection{The Neutrino-Nucleon Cross Section in the CDP.}

Finally, I come back to the neutrino-nucleon cross section to be predicted at
ultrahigh energies. Inserting the expression for $F^{ep}_2 (x, Q^2)$, or rather
the photoabsorption cross section (\ref{2.21}) of the CDP, into the expression
for the
neutrino-nucleon cross section (\ref{2.1}), we find \cite{PRD88}
\bqa
\sigma_{\nu N}(E) &=& \frac{G^2_F M^4_W}{8 \pi^3 \alpha} 
\frac{n_f}{\sum_q Q^2_q} \int^{s-M_p^2}_{Q^2_{Min}} d Q^2 
\frac{Q^2}{(Q^2 + M^2_W)^2}
\nonumber \\
&\times& \int^{s-Q^2}_{M_p^2} \frac{dW^2}{W^2} \frac{1}{2} (1 + (1-y)^2) 
\sigma_{\gamma^*p} (\eta (W^2, Q^2)).
\label{2.25}
\eqa
Upon numerical evaluation of (\ref{2.25}), based on the photoabsorption cross
section from Section 2.5, we find the results shown in fig. 7 \cite{PRD88}. 
\begin{figure}[h!]
\begin{center}
\includegraphics[scale=0.55]{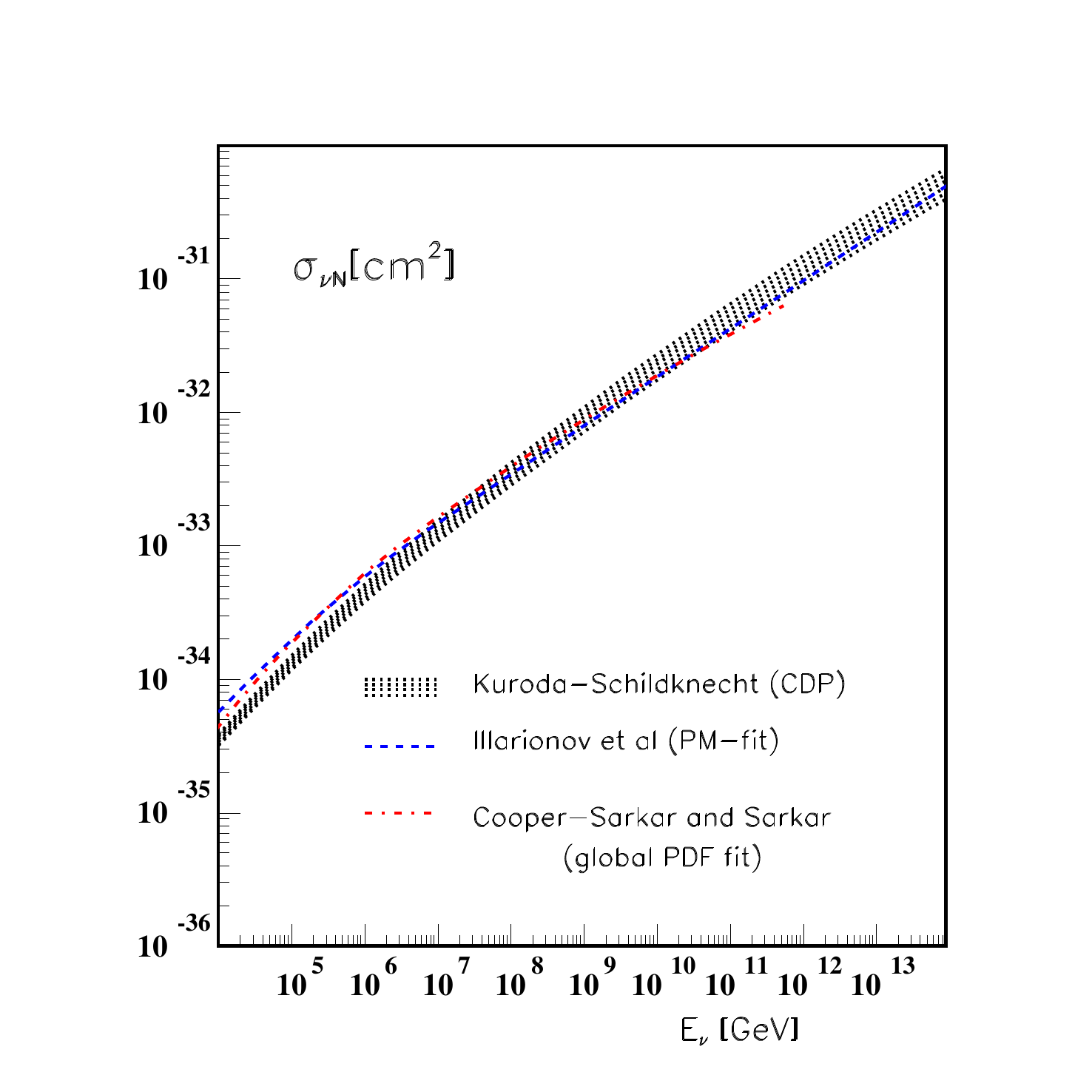}
\vspace*{-0.5cm}
\caption{The prediction \cite{PRD88} of the neutrino-nucleon cross section from the
CDP. For comparison, we also show the results based on the pQCD improved
parton model fit, the ``global PDF fit'', to the HERA experimental data.
The PM-fit is a convenient parameterization of the global PDF fit.}
\end{center}
\vspace*{-0.3cm}
\end{figure}

The shaded area in fig. 7 corresponds to a change of the exponent $C_2$ between 
$C_2 = 0.27$ and $C_2 = 0.29$. In fig. 7, we see that our prediction, based on
a small number of parameters, essentially $C_1$ and $C_2$ to determine
$\Lambda^2_{sat} (W^2)$ supplemented by the bounds
$m^2_0$ and $m^2_1 (W^2) = \xi \Lambda^2_{sat} (W^2)$ on the
actively contributing $q \bar q$ dipole states, coincide with the
very elaborate pQCD fits to the HERA data, the global PDF fit, 
obtained by the big pQCD
collaborations
with a huge number of fairly arbitrary fit parameters. The results of the PM
fit in fig. 6 are based on a six-arbitrary-parameter 
fit to the global PDF fit for the
parton distribution functions at low x.

\subsection{Comparison with the Froissart-inspired Ansatz.}

As early as in 1953, Heisenberg \cite{Heisenberg} considered the question
of the behavior of cross sections among strongly-interacting particles
at asymptotically high energies. Picturing the proton as a Lorentz-contracted
sphere with exponentially decreasing edge, and requiring a minimum
blackness necessary for particle production to occur, he predicted an increase
of strong-interaction cross sections with energy as $(\ln (W^2))^2$. In 
1961, it was shown by Froissart \cite{Froissart} that a growth as
$(\ln W^2)^2$ is the maximum growth allowed by unitarity and
analyticity in quantum field theory.
\begin{figure}[h!]
\begin{center}
\includegraphics[scale=0.5]{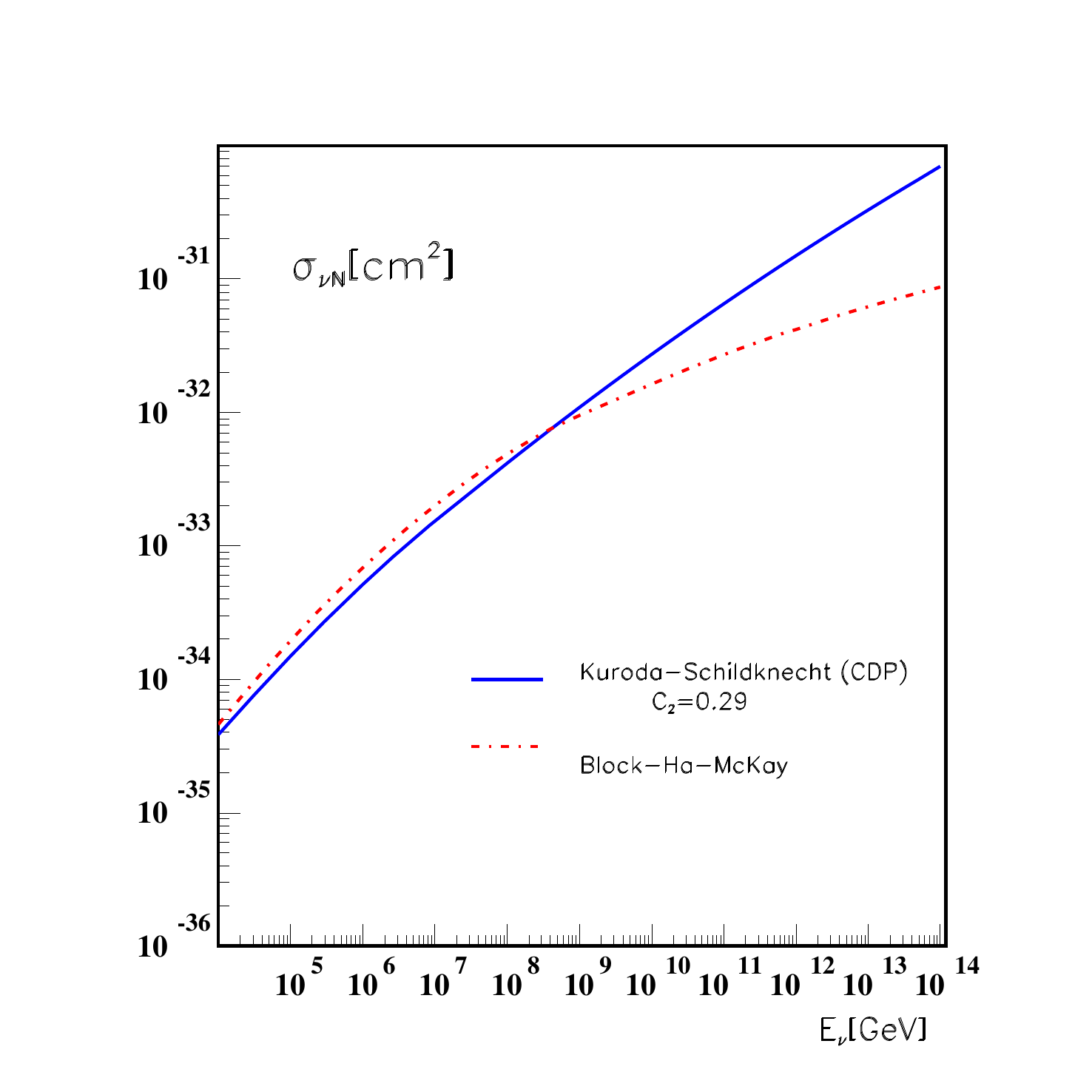}
\vspace*{-0.3cm}
\caption{Comparison \cite{PRD88} of the predictions from the CDP with the
results from the ``Froissart-inspired'' ansatz.}
\end{center}
\vspace*{-0.8cm}
\end{figure}
\begin{figure}[h!]
\begin{center}
\includegraphics[scale=0.5]{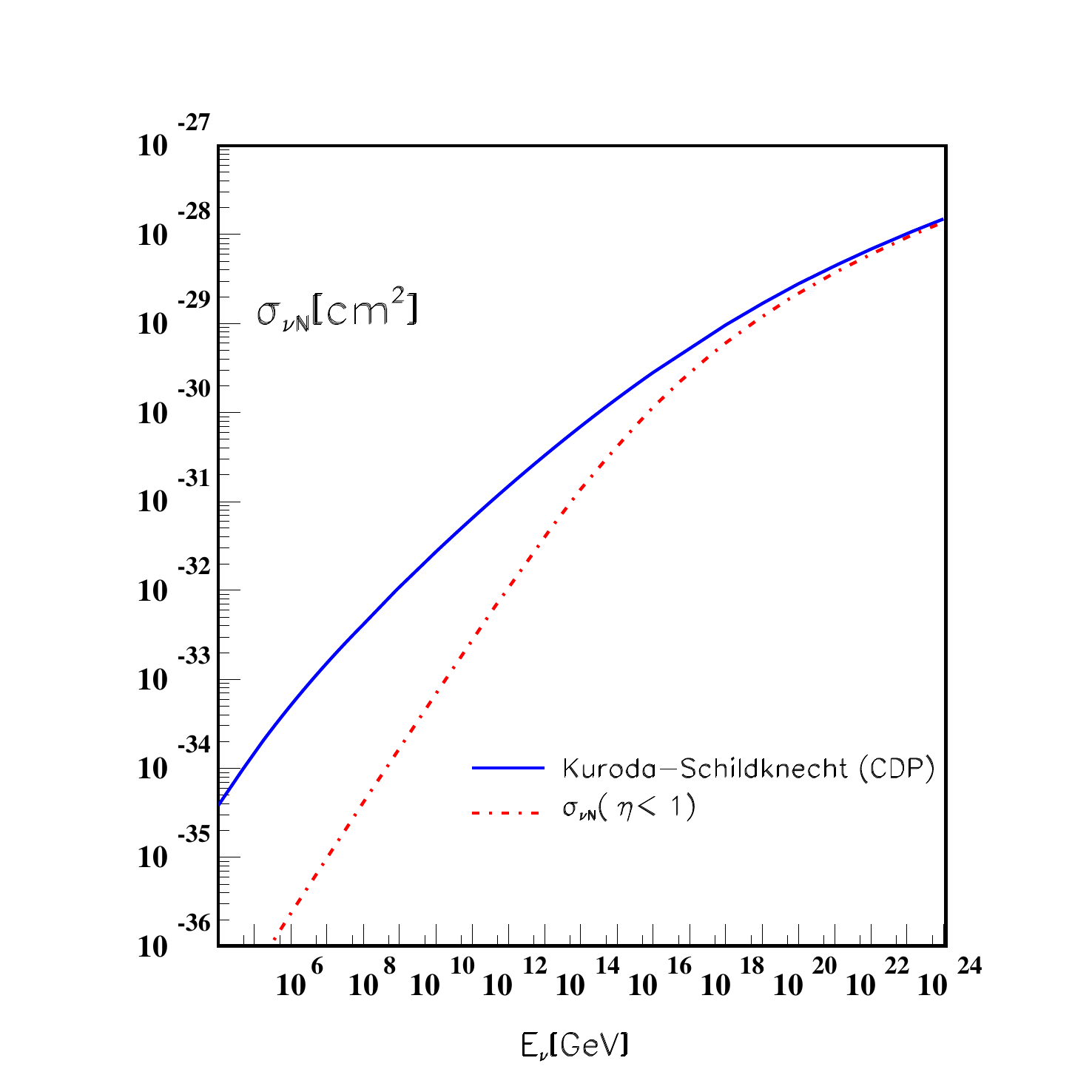}
\vspace*{-0.3cm}
\caption{The neutrino-nucleon cross section compared \cite{PRD88} with the part of
this cross section that is due to the saturation region characterized
by the kinematic constraint of $\eta (W^2,Q^2) < 1$.}
\end{center}
\vspace*{-0.8cm}
\end{figure}

In a series of papers by Block et al \cite{Block} - \cite{Durand}, it was
recently shown that a hadron-like ``Froissart-inspired'' ansatz of the 
form $F^{ep}_2 (x, Q^2) \sim \sum_{n,m = 0,1,2} a_{nm} (\ln Q^2)^n
(\ln (1/x))^m$, with seven free fit parameters provides an excellent
representation of the low-x DIS data from HERA. The prediction for
ultrahigh-energy neutrino scattering \cite{Block-Ha}, based on employing in
(\ref{2.1}) with (\ref{2.6}) the Froissart-inspired ansatz, is shown in
fig. 8. There is satisfactory agreement with the predictions from the 
CDP (and accordingly with the predictions from the pQCD-improved
parton model) for energies up to $E \lsim 10^8 {\rm GeV} = 100 {\rm PeV}$, two
orders of magnitude above the energy of the recently observed IceCube events.

The Froissart-inspired ansatz, according to fig. 8, for energies
$E \gsim 10^8 {\rm GeV}$ implies a weaker growth with energy than the 
prediction from the CDP. The CDP, according to (\ref{2.21}) with
(\ref{2.23}) and $\sigma_{\gamma p}^{PDG} (W^2)$ in (\ref{2.24}), in
the saturation region of $\eta (W^2,Q^2) \ll 1$ contains the same
Froissart-like $(\ln W^2)^2$ behavior as the Froissart-inspired approach.
The differences in fig. 8 must accordingly be due to the differences in the
$Q^2$ dependences between these two approaches. According to the CDP,
as shown in fig. 9, at energies even far beyond $E \cong 10^8 {\rm GeV}$,
for energies below $E \cong 10^{12}  {\rm GeV}$,
the contribution to the neutrino-nucleon cross section from the
saturation region of $\eta (W^2,Q^2) < 1$  is suppressed by at least one order of
magnitude. Contributions from the $(\ln W^2)^2$ behavior in the saturation
region only start to dominate the neutrino-nucleon cross section at
unrealistic ultra-ultrahigh energies beyond the upper end
of $E \cong 10^{12} {\rm GeV}$
  of the cosmic-ray spectrum in fig. 1.

Our results from the CDP differ significantly from various
results in the literature incorporating non-linear evolution effects
\cite{Balitzki} in the pQCD improved parton-model-approach. The results
imply \cite{Kutak, Reno, Goncalves, Machado} neutrino cross sections that
at $E \cong 10^{12} {\rm GeV}$ are approximately a factor of two to three
below our predictions in figs. 7 and 8.

\section{Conclusions}

Inspired by cosmic-neutrino-search experiments, we thoroughly examined
the neutrino-nucleon cross section at ultrahigh energies. Our results,
based on the simple closed analytic expression for the photoabsorption
cross section at low x in the CDP, in the full range of energies
$E \lsim 10^{12} {\rm GeV}$, are consistent with the results based on
the extrapolation of the elaborate multi-parameter global pdf fits
to the HERA experimental data.

Below energies of $E \lsim 10^8 {\rm GeV}$, two orders of magnitude
above the energy $E \approx 10^6 {\rm GeV}$ of the recently detected
IceCube events, our CDP results also agree with the predictions from the
Froissart-inspired approach to DIS.

Above energies of $E \gsim 10^8 {\rm GeV}$, the predictions from the CDP
disagree with both, the
predictions from the Froissart-inspired approach, and the
predictions incorporating non-linear evolution of parton-distribution
functions. These predictions lead to a suppressed cross section, when
compared with the CDP, the suppression reaching factors of
approximately three at $E \approx 10^{12} {\rm GeV}$.

The CDP-based analysis shows that the ultrahigh-energy neutrino-nucleon
cross section, in the full range of energies, up to 
$E \approx 10^{12} {\rm GeV}$,
of the cosmic-ray spectrum, is dominated by contributions from
the kinematic region, where color transparency determines the interaction.
The dominance of color transparency is responsible for the consistency
of the CDP neutrino-nucleon cross section with the global pdf fits,
while excluding the presence of screening effects due to non-linear
evolution.

\section*{Acknowledgement}

The author thanks Kuroda-san for fruitful collaboration on the color
dipole picture.

It is a great pleasure for the author to thank
Professor Antonino Zichichi for the invitation to the
52nd Course of the International School of Subnuclear Physics with the lively
scientific atmosphere in the magnificent environment of the Ettore
Majorana Foundation and Centre for Scientific Culture at Erice, Sicily.

\end{document}